\documentclass[USenglish,twocolumn]{article}
\usepackage[utf8]{inputenc}
\usepackage[big,online]{dgruyter}
\usepackage[sort&compress,square,numbers]{natbib}
\usepackage{times}
\usepackage{graphicx}
\usepackage{subfig}
\usepackage{placeins}
\usepackage[separate-uncertainty  = true,
uncertainty-separator =  {\,},
output-decimal-marker = {.},
multi-part-units = single,
range-units = brackets,
range-phrase = {\,--\,}]{siunitx}
\usepackage{tikz}
\usepackage{xcolor}
\usepackage{pgfplots}
\usepackage{amsmath}
\usetikzlibrary{shapes,arrows.meta,3d,calc,fit,shadows,snakes,automata,backgrounds,petri,positioning}
\begin{document}

  \DOI{10.1515/}
  \openaccess
  \pagenumbering{gobble}

\title{Do We Need Pre-Processing for Deep Learning Based Ultrasound Shear Wave Elastography?}

\runningtitle{Pre-Processing in Deep Learning Based US-SWE}

\author*[1]{Sarah Grube}
\author[2]{Sören Grünhagen}
\author[2]{Sarah Latus} 
\author[2]{Michael Meyling} 
\author[3]{Alexander Schlaefer} 
\runningauthor{S.~Grube et al.}

\affil[1]{\protect\raggedright 
 Hamburg University of Technology, Institute of Medical Technology and Intelligent Systems, Hamburg, Germany, e-mail: sarah.grube@tuhh.de}
\affil[2]{\protect\raggedright
  Hamburg University of Technology, Institute of Medical Technology and Intelligent Systems, Hamburg, Germany}
\affil[3]{\protect\raggedright
   Hamburg University of Technology, Institute of Medical Technology and Intelligent Systems, Hamburg, Germany and SustAInLivWork Center of Excellence}

\abstract{
Estimating the elasticity of soft tissue can provide useful information for various diagnostic applications. Ultrasound shear wave elastography offers a non-invasive approach. However, its generalizability and standardization across different systems and processing pipelines remain limited. 
Considering the influence of image processing on ultrasound based diagnostics, recent literature has discussed the impact of different image processing steps on reliable and reproducible elasticity analysis. 
In this work, we investigate the need of ultrasound pre-processing steps for deep learning-based ultrasound shear wave elastography. 
We evaluate the performance of a 3D convolutional neural network in predicting shear wave velocities from spatio-temporal ultrasound images, studying different degrees of pre-processing on the input images, ranging from fully beamformed and filtered ultrasound images to raw radiofrequency data. We compare the predictions from our deep learning approach to a conventional time-of-flight method across four gelatin phantoms with different elasticity levels. 
Our results demonstrate statistically significant differences in the predicted shear wave velocity among all elasticity groups, regardless of the degree of pre-processing.
Although pre-processing slightly improves performance metrics, our results show that the deep learning approach can reliably differentiate between elasticity groups using raw, unprocessed radiofrequency data.
These results show that deep \mbox{learning-based} approaches could reduce the need for and the bias of traditional ultrasound pre-processing steps in ultrasound shear wave elastography, enabling faster and more reliable clinical elasticity assessments.}

\keywords{Raw radiofrequency data, Shear wave elastography, Ultrasound, Convolutional neural network, Pre-Processing}

\maketitle

\section{Introduction} 
The mechanical properties of tissues can provide diagnostic information and help in the diagnosis and treatment of diseases. For example, tumors often have increased stiffness compared to surrounding healthy tissue~\cite{chan2021}. Therefore, differences in tissue elasticity can indicate pathological changes.
Ultrasound shear wave elastography (US-SWE) provides a method for non-invasive, real-time assessment of tissue stiffness. This technique is already used in clinical practice, for instance in the staging of liver fibrosis~\cite{ferraioli2014}. In US-SWE, shear waves are induced within the tissue and their propagation is tracked. The shear wave propagation velocity is directly related to the Young's modulus and provides a quantitative estimation of tissue stiffness~\cite{Sarvazyan1995}. 
However, the accuracy of US-SWE is influenced by the system configurations and processing algorithms applied to the acquired raw US data~\cite{nitta2021}. 
Therefore, the generalizability and standardization of US-SWE remains an important goal for further clinical application. 
Typical US-SWE processing steps can be divided into B-mode image reconstruction, shear wave-specific filtering, such as the Loupas filter~\cite{loupas1995}, and velocity estimation.
Various algorithms have been proposed for these steps, each of which introduces methodological assumptions and potential sources of inaccuracy.
\mbox{B-mode} image reconstruction, for instance, generally assumes a constant speed of sound. However, biological tissue is heterogeneous, and therefore the assumption of uniform speed of sound can lead to significant spatial distortions~\cite{bland2017, nitta2021}. 
Furthermore, research has demonstrated that the accuracy of US-SWE also depends on the processing algorithms and their selected parameters.
The parameter settings used in shear wave tracking algorithms, such as configurations of the directional filters and kernel dimensions, have a significant impact on measurement results ~\cite{deng2016, rouze2012, nitta2021}.
Deng et al. have shown that the US-SWE measurement errors can be significantly reduced by carefully optimizing these parameters for a particular system configuration~\cite{deng2016}. However, such tuning is usually system-specific, which limits the ability to generalize to different platforms.

In order to improve the reliability and reproducibility of US-based measurements, recent literature has proposed minimizing processing steps. 
For example, Sanabria et al. showed that combining raw beamformed data with deep learning approaches can improve tissue quantification~\cite{Sanabria2022}. 
To summarize, current literature indicates that raw or minimally processed US data tends to retain more diagnostic information compared to highly processed B-mode images. Therefore, reducing the number of processing steps can increase robustness, minimize bias, and reduce information loss caused by processing algorithms.

In this study, we investigate whether deep learning approaches can estimate shear wave velocities from minimally pre-processed US data. Starting with highly processed US data from a conventional US-SWE processing pipeline, we systematically remove processing steps. Finally, we evaluate the feasibility of predicting shear wave velocity directly from raw radiofrequency~(RF) signals. To the best of our knowledge, we are the first to systematically analyze the effect of ultrasound data pre-processing on deep learning-based shear wave velocity estimation, particularly considering predictions directly from raw RF data without any beamforming.

\section{Methods} 
\subsection{Data Acquisition and Processing}
The data acquisition is described in detail in \cite{grube2022, grube2023}. Shear waves were induced in four gelatin phantoms~(\SI{15}{\%}, \SI{12.5}{\%}, \SI{10}{\%}, and \SI{7.5}{\%} gelatin) using a piezo-actuated needle mounted on a UR5~(Universal Robots, DK) robotic arm. A linear ultrasound probe~(\text{Ultrasonix L14-5/38}) mounted on a UR3 robot~(Universal Robots, DK) tracked the resulting shear wave propagation over time using plane wave imaging.
A 256 channel US system (Griffin, Cephasonics, USA) and SUPRA~\cite{Göbl2018} were used to acquire the raw RF US data with an image acquisition rate of 6000 Hz. For each acquisition, 70 frames were acquired.
To increase variability in shear wave velocities, 240 measurements for each phantom were repeated on three consecutive days, with a slight increase in velocity observed over time due to gelatin hardening~\cite{grube2024}.
The resulting data set consists of 12 different stiffness values, corresponding to shear wave velocities ranging from \SI{3.77}{m/s} to \SI{8.42}{m/s}.

\begin{figure}
	\includegraphics[width=\linewidth]{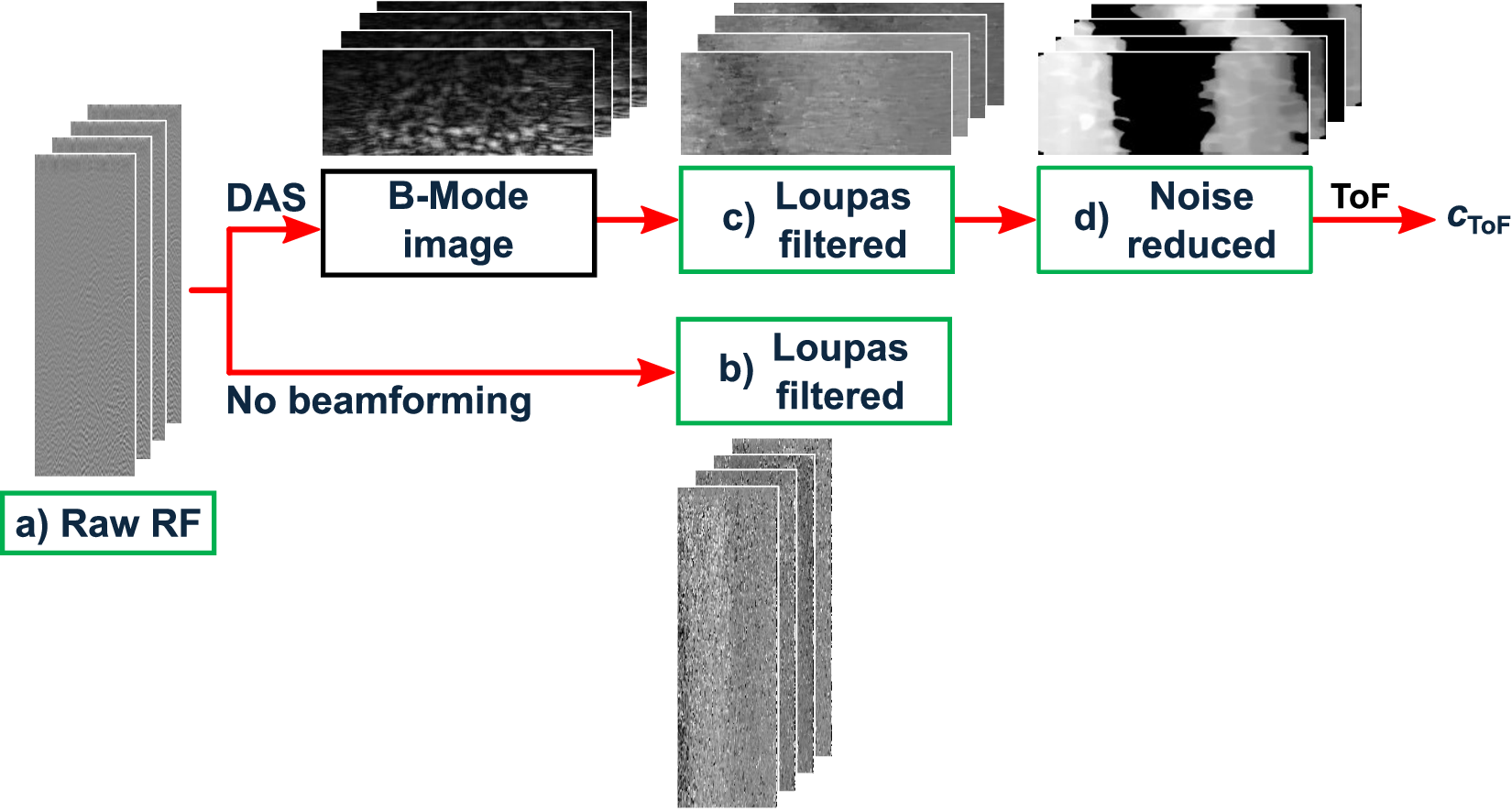}
	\caption{US-SWE pre-processing pipeline: the US images after the different processing steps are shown. Raw RF-data a), reconstructed B-mode images, Loupas-filtered B-Mode US images c) and additional filtering d) for noise reduction. Loupas filtering directly on raw RF data is also considered b). Data types used for training our network are marked in green.}
	\label{fig:dataProcessingPipeline}
\end{figure}

The raw RF data were processed according to the pre-processing pipeline used in Grube~et~al.~\cite{grube2023}, which is shown schematically in Figure ~\ref{fig:dataProcessingPipeline}. Delay-and-sum~(DAS) beamforming was first applied to transform the raw RF data into B-mode images. Then, the Loupas autocorrelation algorithm~\cite{loupas1995} was applied to visualize the shear wave in the image. For better shear wave visibility, noise was reduced, using a combination of median filtering and morphological operations. In the next step, a time-of-flight~(ToF) method was used to determine the shear wave velocity $c_{\text{ToF}}$.

\subsection{Neural Network Architecture}
\begin{figure}
	\includegraphics[width=\columnwidth]{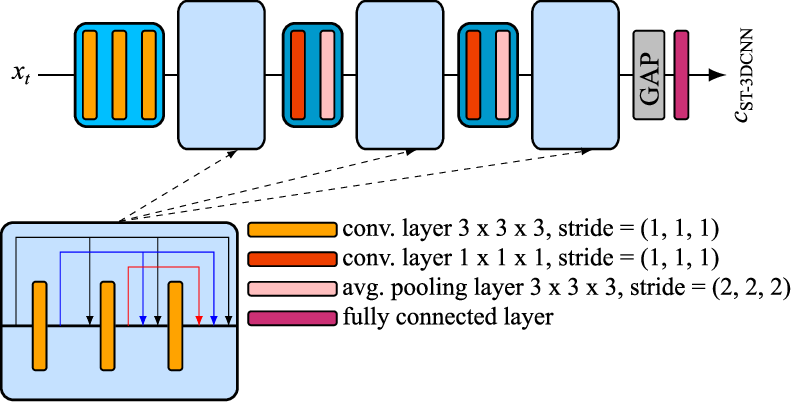}
	\caption{Schematic drawing of the \mbox{ST-3DCNN} used to estimate the shear wave velocity $c_{\text{ST-3DCNN}}$ based on a 3D input sequence $x_t$.}
	\label{fig:st3dcnn}
\end{figure}

The proposed spatio-temporal convolutional neural network~(ST-3DCNN) is based on our previous work~\cite{grube2023} and is shown schematically in Figure~\ref{fig:st3dcnn}. 
The network estimates shear wave velocity from a time-sequence of 2D US images by solving a regression task. 
Each input image sequence~$x_t$ is a spatio-temporal tensor of size \( h \times w \times t \), where \( h \) and \( w \) denote the height and width of the US images, and \( t \) is the number of frames over time.
The regression task learns a function \( f: \mathbb{R}^{h \times w \times t} \rightarrow \mathbb{R} \), which maps the input image sequence to a scalar prediction of the shear wave velocity $c_{\text{ST-3DCNN}}$. 
Our \mbox{ST-3DCNN} consists of three initial convolutional layers with five feature maps, followed by three DenseNet blocks with a growth rate of five. Between the blocks, transition layers with convolution and average pooling are used for downsampling.

\subsection{Experiments}
We investigate the performance of our network on the four differently processed data sets, marked in green in Figure~\ref{fig:dataProcessingPipeline}.
Specifically, we focus on the effects of omitting the noise reduction filter, the Loupas filter, and the beamforming algorithm.\\
We train our architecture with a mean squared error (MSE) loss function between the estimated shear wave velocity $c_{\text{ST-3DCNN}}$ and the training label that was measured with ToF. We use the Adam optimizer with a learning rate of 0.001 and a batch size of 10, and train our networks for 150 epochs using four-fold cross-validation. 
For training and validation, we use \SI{70}{\%} and \SI{30}{\%} of the data, respectively, using the first two days of data acquisition. For testing, we use the data acquired on the third day, independent of our training and validation set.
To augment the dataset, each image sequence was split into five non-overlapping subsequences, each consisting of 12 frames over time. 
This resulted in 6720, 2880, and 4800 samples used for training, validation and testing, respectively.\\
Only the raw RF data was used for training in its original resolution. The further processed data was downsampled by a factor of 8 to reduce the training time. Please note that we performed a comparison between models trained on original versus downsampled Loupas data showing no significant performance differences, confirming downsampling as a valid approach.
Please note that during network training, we assign the same label to all input data within each elasticity group, using the mean ToF-derived velocity for each elasticity group as the training label.
During evaluation of the test dataset, however, we compare each predicted shear wave velocity from the \mbox{ST-3DCNN} to the corresponding ToF velocity of the specific input image sequence. This accounts for variability in speed, which is also present in ToF measurements.

\section{Results and Discussion}
\begin{figure}[tb]
	\centering
    \subfloat[raw RF data]{
        {\includegraphics[width=0.45\linewidth]{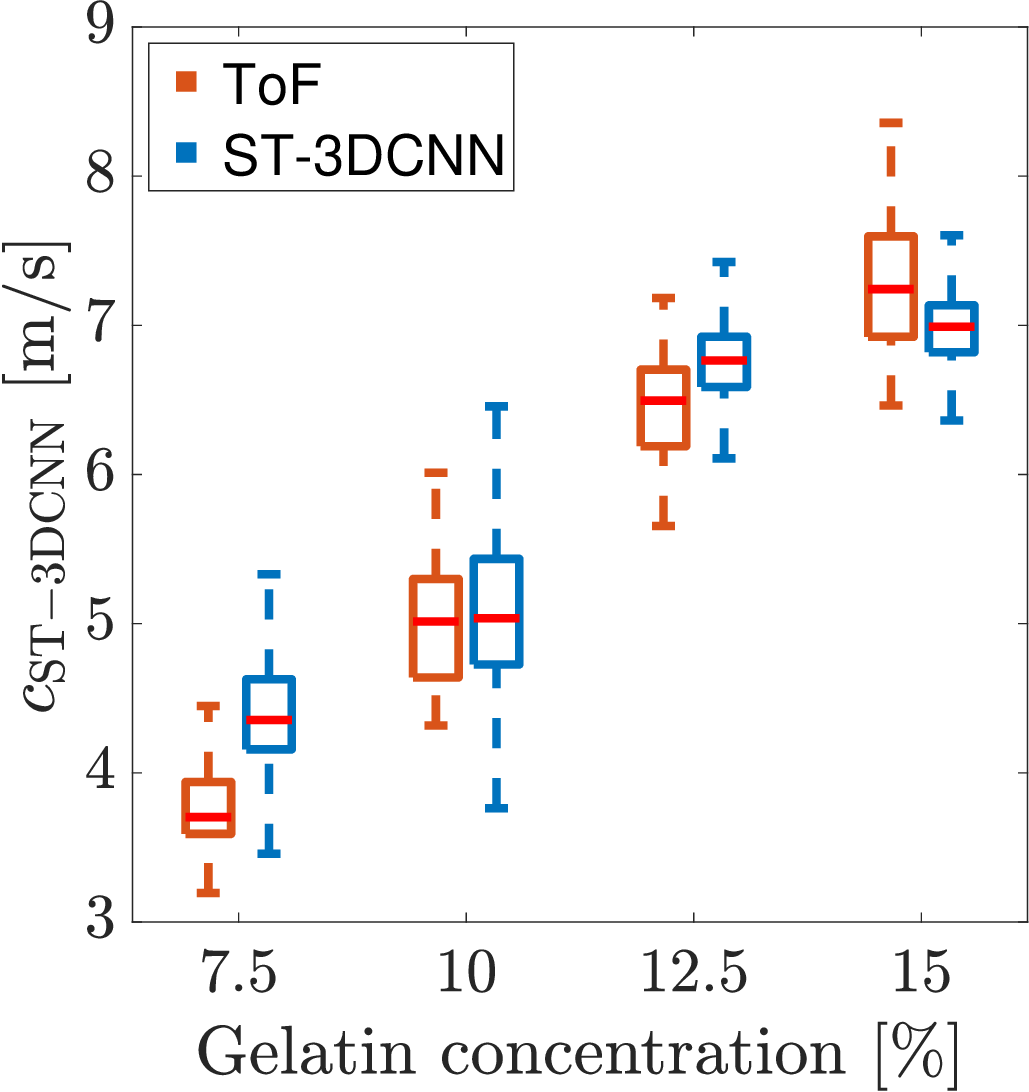}}
        \label{fig:generalLabSetup}
    }
    \hfill
    \subfloat[raw RF data + Loupas]{
        \includegraphics[width=0.45
        \linewidth]{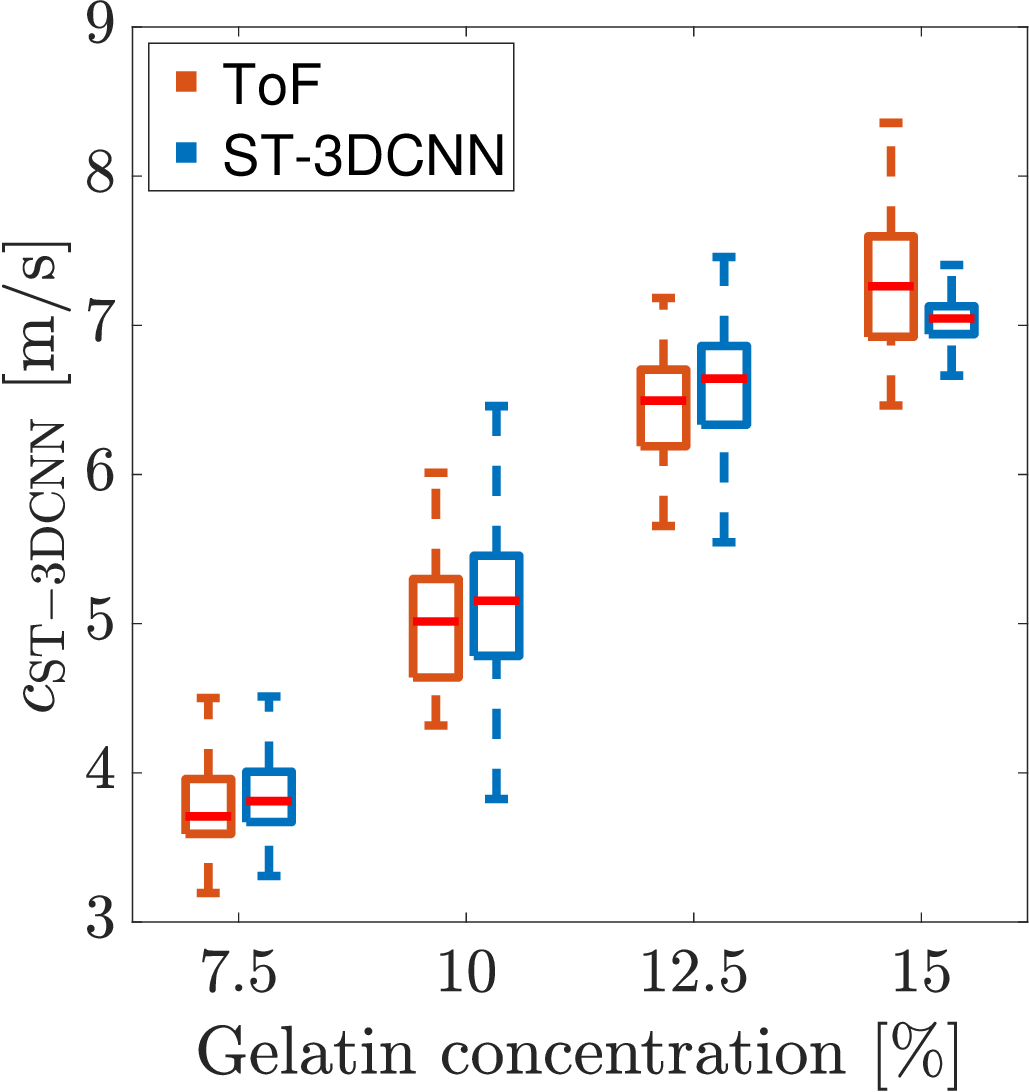}
        \label{fig:GeneralWorkflow_hochkant}
    }
    \hfill
    \subfloat[DAS + Loupas]{
        {\includegraphics[width=0.45\linewidth]{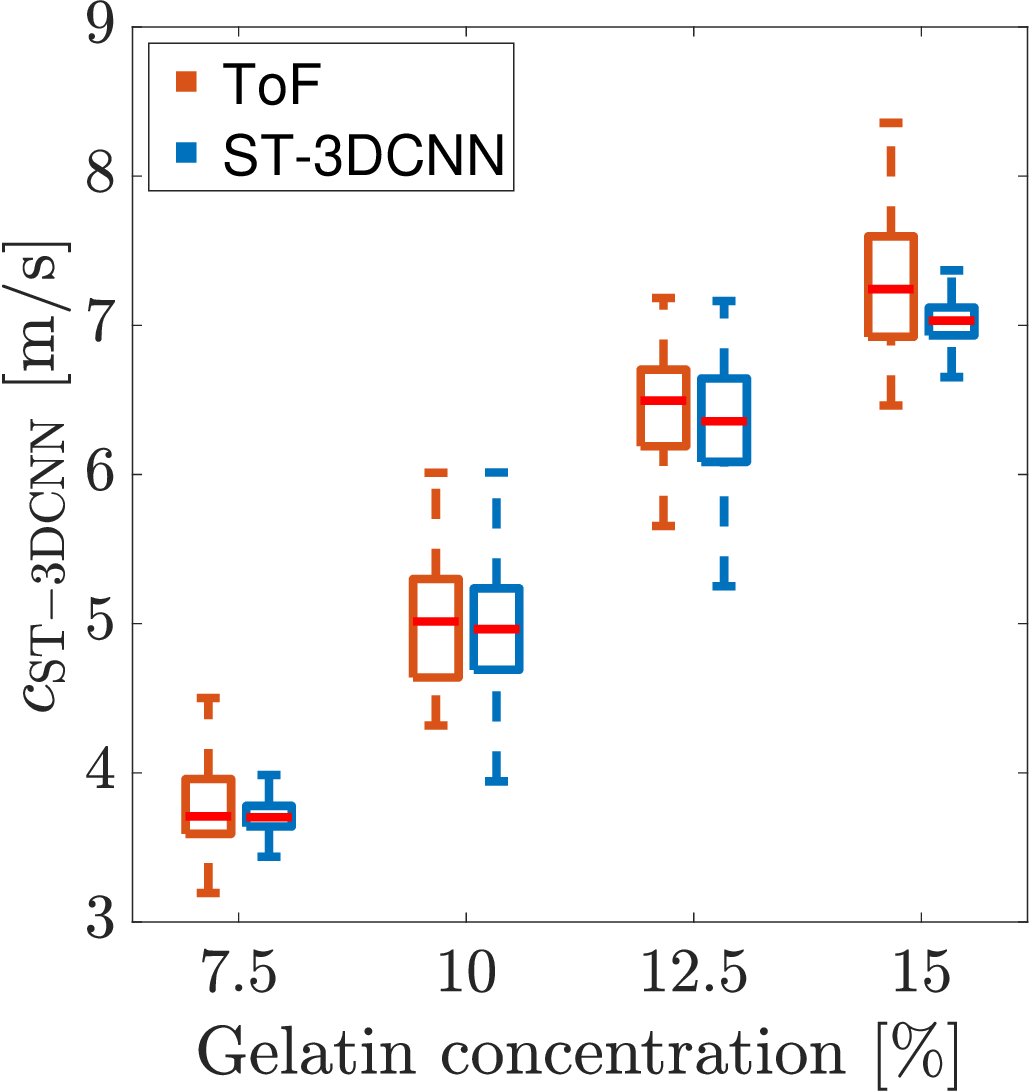}}
        \label{fig:generalLabSetup}
    }
    \hfill
    \subfloat[DAS + Loupas + denoising]{
        {\includegraphics[width=0.45\linewidth]{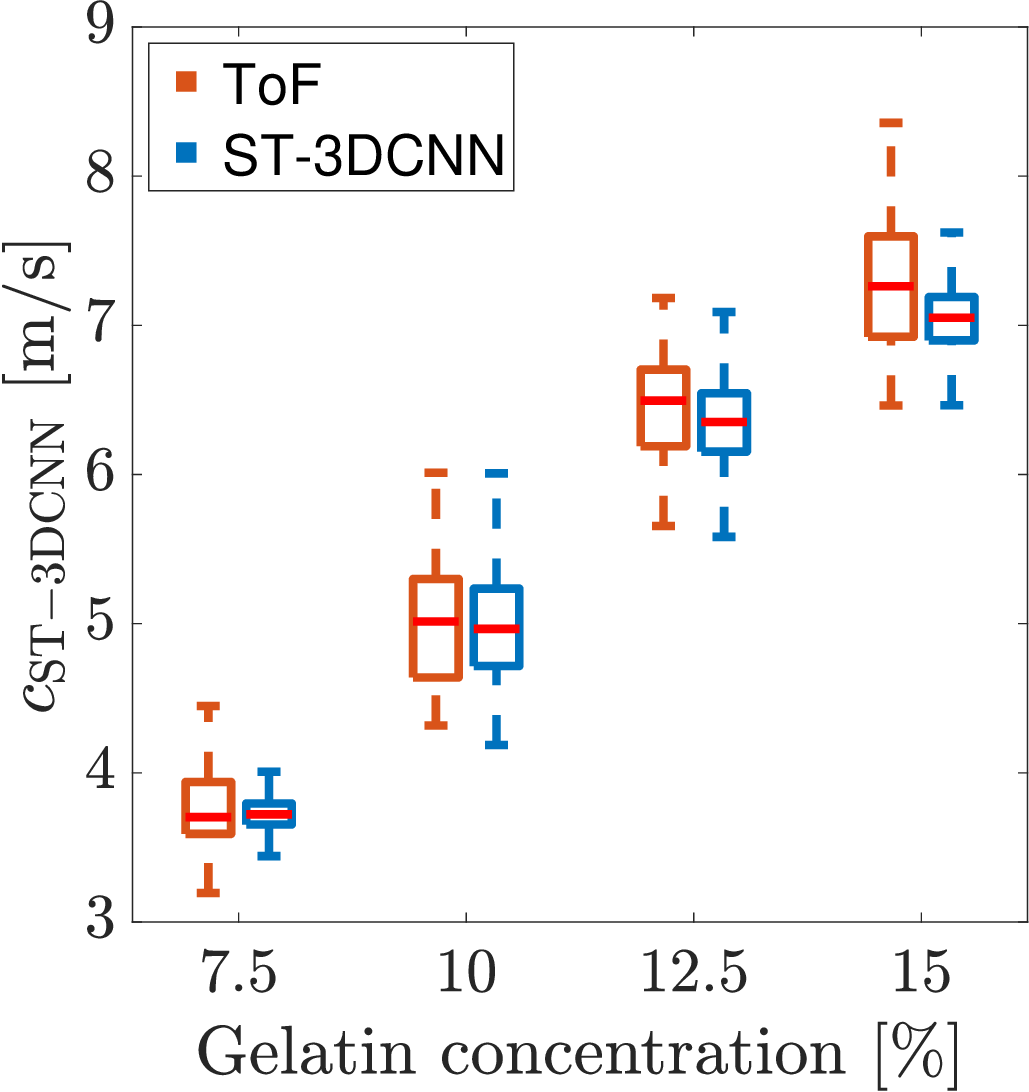}}
        \label{fig:generalLabSetup}
    }
	\caption{Boxplots of the estimated shear wave velocities under different processing methods. (a) No beamforming, using raw RF data; (b) no beamforming with Loupas filtering applied directly to raw RF data; (c) beamforming with Loupas filtering; and (d) beamforming with Loupas filter and additional noise reduction. Note, that for reference the boxplots for the $c_{\text{ToF}}$ are also shown.}	
	\label{fig:tofVSdnn}
\end{figure}

Table~\ref{tab:tof_mean} reports the mean shear wave velocity and standard deviation obtained using the ToF method for each gelatin phantom. Results from the \mbox{ST-3DCNN} approach are summarized in Table~\ref{tab:st3dcnn} and visualized as boxplots alongside the ToF results in Figure~\ref{fig:tofVSdnn}.
We test for significant differences in the median of the mean absolute error~(MAE) of our methods using the Wilcoxon signed-rank test with a significance level of $\alpha$~=~\SI{5}{\%}. 
Outliers were defined as data points differing by more than 1.5 interquartile ranges from the median. 
Statistically significant differences were observed between the predicted values for the four elasticity groups across all investigated data types. This demonstrates that all investigated data types can be used to distinguish the phantoms by their stiffness levels.

The MAE and standard deviation between the shear wave velocities predicted by the \mbox{ST-3DCNN} and those estimated using the conventional ToF method decreased with each additional pre-processing step. The MAE decreased from \SI{0.52 \pm 0.38}{m/s} with raw RF data without beamforming to \SI{0.26 \pm 0.29}{m/s} with fully beamformed and filtered input data. Concurrently, the coefficient of determination~($R^2$) between the ToF-derived velocities and the \mbox{ST-3DCNN} predictions increased from $0.79$ to $0.92$ with more advanced processing.
This performance improvement can be explained by better noise suppression and data homogenization through pre-processing, as well as the growing similarity between the input image sequence and the training label, which was based on the ToF method.
It should be noted, however, that the network cannot outperform the biased ToF based reference it was trained on. 
Hence, in future experiments training labels independent of the ToF method should be used to achieve better performance on less processed data.

Furthermore, the \mbox{ST-3DCNN} resulted in fewer outliers than the ToF approach, especially in the lowest elasticity group. This indicates that the network is more robust to noise and can generalize better in challenging signal conditions compared to the ToF method.

\begin{table}
\centering
\caption{Mean shear wave velocity and standard deviation (std) in [m/s] using the ToF method, along with the number of outliers per elasticity group (\SI{7.5}{\%} -- \SI{15}{\%} gelatin concentration).}
\begin{tabular}{lll}
Elasticity group 	& $c_{\text{ToF}}$ $\pm$ std	& Outliers~[\%]		\\ \midrule
\SI{7.5}{\%} 	&  	\SI{3.90 \pm 0.57}{m/s}   &\num{18.58}\\
\SI{10}{\%}  & 	\SI{5.00 \pm 0.38}{m/s}   &\num{1.69}\\
\SI{12.5}{\%}  & 	\SI{6.44 \pm 0.34}{m/s}   &\num{0} \\
\SI{15}{\%} & \SI{7.29 \pm 0.39}{m/s}  &\num{0.45} \\
\end{tabular}
\label{tab:tof_mean}
\end{table}

\begin{table}
\centering
\caption{Mean absolute error (MAE) and standard deviation (std) of the predicted shear wave velocities $c_{\text{ST-3DCNN}}$ relative to the corresponding ToF velocity. 
Results for the four processing methods investigated are shown, averaged over all elasticity groups: (a) no beamforming, using raw RF data; (b) no beamforming with Loupas filtering applied directly to raw RF data; (c) beamforming with Loupas filtering; and (d) beamforming with Loupas filter and additional noise reduction. The $R^2$ coefficient between the ToF-based and \mbox{ST-3DCNN}-based velocity estimates is also reported, along with the number of outliers.}
\begin{tabular}{llll}
Data type 	& MAE $\pm$ std &$R^2$	&Outliers [\%]		\\ \midrule
(a) 	&  	\SI{0.52 \pm 0.38}{m/s} &\num{0.79} & $1.40$\\
(b)  & \SI{0.43 \pm 0.37}{m/s}	&$0.82$ &$2.19$\\
\midrule
(c)  & \SI{0.40 \pm 0.36}{m/s} &$0.85$ &$1.90$\\
(d) &\SI{0.26 \pm 0.29}{m/s} &$0.92$ &$0.88$\\
\end{tabular}
\label{tab:st3dcnn}
\end{table}

\section{Conclusion} 
US-SWE shows promising potential to support many clinical applications. In this work, we investigated the need for pre-processing in deep learning-based US-SWE.
Our study was motivated by the benefits of using raw US data, in particular the preservation of information that might otherwise be lost through conventional pre-processing. 
We progressively reduced the number of processing steps applied to the input image sequence, from conventionally processed signals to raw RF data, and estimated shear wave velocities using a spatio-temporal convolutional neural network.
Our results demonstrate that tissue stiffness can be successfully distinguished using deep learning methods even without any pre-processing, highlighting the feasibility of operating directly on raw US data without the need for beamforming. In future work, network architectures specifically optimized for raw RF data could be investigated in more detail.

\textsf{\textbf{Author Statement}}\\
Research funding: This work was partially funded by the TUHH i3 initiative, the Interdisciplinary Competence Center for Interface Research (ICCIR) supported by TUHH and UKE, and by the Deutsche Forschungsgemeinschaft under grant SCHL 1844/6-1. In addition, this research was co-funded by the European Union under Horizon Europe program (No. 101059903); and by the European Union funds for the period 2021--2027. 
Conflict of interest: Authors state no conflict of interest. Informed consent: Informed consent has been obtained from all individuals included in this study. Ethical approval: No ethical approval was necessary for this research.

\begin{thebibliography}{99}
\bibitem{bland2017}
Bland T, Tong J, Ward B, Parker NG. Geometric distortion of area in medical ultrasound images. J Phys Conf Ser 2017;797(1):012002. 

\bibitem{chan2021}
Chan HW, Uff C, Chakraborty A, Dorward N, Bamber JC. Clinical application of shear wave elastography for assisting brain tumor resection. Front Oncol 2021;11:619286.

\bibitem[4]{deng2016}
Deng Y, Rouze NC, Palmeri ML, Nightingale KR. On system-dependent sources of uncertainty and bias in ultrasonic quantitative shear-wave imaging. IEEE T-UFFC 2016;63(3):381-393.                   

\bibitem{ferraioli2014}
Ferraioli G, Parekh P, Levitov AB, Filice C. Shear wave elastography for evaluation of liver fibrosis. J Ultrasound Med 2014;33(2):197-203.

\bibitem{grube2023}
Grube S, Bengs M, Neidhardt M, Latus S, Schlaefer A. Ultrasound shear wave velocity estimation in a small field of view via spatio-temporal deep learning. Med Imaging 2023: Image Processing. SPIE 2023;504-508.  

\bibitem[6]{grube2024}
Grube S, Neidhardt M, Hermann A, Sprenger J, Abdolazizi K, Latus S, et al. A calibration approach for elasticity estimation with medical tools. Curr Dir Biomed Eng 2024;10(2):99-102.

\bibitem{grube2022}
Grube S, Neidhardt M, Latus S, Schlaefer A. Influence of the field of view on shear wave velocity estimation. Curr Dir Biomed Eng 2022;8(1):42-45.
     
\bibitem{Göbl2018}
Göbl R, Navab N, Hennersperger C. SUPRA: Open source software defined ultrasound processing for real-time applications. IJCARS 2018;13:759–767.

\bibitem{loupas1995}
Loupas T, Powers JT, Gill RW. An axial velocity estimator for ultrasound blood flow imaging, based on a full evaluation of the Doppler equation by means of a two-dimensional autocorrelation approach. IEEE T-UFFC 1995;42(4):672-688.

\bibitem[5]{nitta2021}
Nitta N, Yamakawa M, Hachiya H, Shiina T. A review of physical and engineering factors potentially affecting shear wave elastography. J Med Ultrasonics 2018;48:403–414. 

\bibitem[3]{rouze2012}
Rouze NC, Wang MH, Palmeri ML, Nightingale KR. Parameters affecting the resolution and accuracy of 2-D quantitative shear wave images. IEEE T-UFFC 2012;59(8):1729-1740.

\bibitem{Sanabria2022}
Sanabria SJ, Pirmoazen AM, Dahl J, Kamaya A, El Kaffas A. Comparative study of raw ultrasound data representations in deep learning to classify hepatic steatosis. Ultrasound Med Biol 2022;48(10):2060–78.

\bibitem{Sarvazyan1995}
Sarvazyan AP, Skovoroda AR, Emelianov SY, Fowlkes JB, Pipe JG, Adler RS, et al. Biophysical bases of elasticity imaging. In: Jones JP, editor. Acoustical imaging. Vol. 21. Boston, MA: Springer;1995:223–240.

\end{thebibliography}

\end{document}